\documentclass[fleqn,twoside]{article}
\usepackage{espcrc2}

\usepackage{graphicx}
\usepackage[figuresright]{rotating}

\newcommand{\beq}{\begin{equation}}
\newcommand{\eeq}{\end{equation}}
\newcommand{\bea}{\begin{eqnarray}}
\newcommand{\eea}{\end{eqnarray}}

\renewcommand{\l}{\lambda}

\renewcommand{\b}{\beta}

\newcommand{\g}{\gamma}

\newcommand{\m}{\mu}

\newcommand{\s}{\sigma}

\newcommand{\oh}{{\textstyle{\frac{1}{2}}}}

\newcommand{\dg}{\dagger}
\newcommand{\non}{\nonumber}

\newcommand{\rf}[1]{(\ref{#1})}
\newcommand{\ra}{\rightarrow}

\newcommand{\AmS}{{\protect\the\textfont2
  A\kern-.1667em\lower.5ex\hbox{M}\kern-.125emS}}

\title{Remnant Symmetry and the Confinement Phase in
Coulomb Gauge \thanks{Talk given by J. Greensite at {\sl QCD Down Under},
 Adelaide, Australia, March 10-19, 2004.
Research supported in part by the
US Dept.\ of Energy, Grant No.\ DE-FG03-92ER40711 (J.G.), 
the Slovak Grant Agency for Science, Grant No.\ 2/3106/2003 ({\v S}.O), 
and the National Science Foundation, Grant No.\ PHY-0099393 (D.Z.). } }

\author{Jeff Greensite\address{The Niels Bohr Institute,
Blegdamsvej 17, 2100 Copenhagen \O, Denmark}, 
{\v S}tefan Olejn\'{\i}k\address{Institute of Physics, Slovak Academy
of Sciences, SK-845 11 Bratislava, Slovakia}, 
and Daniel Zwanziger\address{Physics Department, New York
University, New York, NY~10003, USA} }

\begin{document}

\begin{abstract}
   We report on connections between the confining
color Coulomb potential, center vortices, and the unbroken realization
of remnant gauge symmetry in Coulomb gauge.
\end{abstract}

\maketitle

\section{The Color Coulomb Potential}

    The color Coulomb potential is derived from the expectation
value of the non-local part of the Coulomb-gauge Hamiltonian, in the
presence of static external charges.  This is a $1/R$ potential in
electrodynamics, but the $R$-dependence may be different in
non-abelian gauge theories.  In this talk I will show that the
confining behavior of the color Coulomb potential depends on the
unbroken realization of a global gauge symmetry, which remains
after imposing Coulomb gauge.  I will also make some connections
to the vortex confinement mechanism, and this theme will be
further developed in Dan Zwanziger's talk. The results outlined
here are reported in much greater detail in ref.\ \cite{US};
I will also touch on the studies in refs.\ \cite{GO} and \cite{Bertle2}.

   Let $|\Psi_{qq}\rangle = \overline{q}(0) q(R) |\Psi_0 \rangle$
denote a physical state in Coulomb gauge containing heavy
quark-antiquark static charges, where $\Psi_0$ is the Yang-Mills vacuum
wavefunctional.  Then
\bea
       {\cal E} &=& \langle \Psi_{qq}|H|\Psi_{qq}\rangle
                - \langle \Psi_0|H|\Psi_0\rangle
\non \\
                &=& V_{coul}(R) + E_{se}
\eea
includes the $R$-dependent color Coulomb potential, plus self-energy
contributions.  It is natural to ask, first, whether $V_{coul}(R)$ is
confining.  If confining, is the potential asymptotically linear?  If
linear, does the Coulomb string tension $\s_{coul}$ equal the usual
string tension $\s$ of the static quark potential?  Finally, what about
center vortices?  What happens to the color Coulomb potential when
vortices are removed?  To address these questions, we define the
correlator of two timelike Wilson (not Polyakov) lines in Coulomb gauge
\bea
    G(R,T) &=& \langle \oh \mbox{Tr}[L^\dg(0,T) L(R,T)] \rangle
\non \\
    V(R,T) &=& -{d \over dT} \log[G(R,T)]
\eea
where
\beq
L(\vec{x},T) = \exp\left[i\int_0^T dt ~ A_0(\vec{x},t)\right]
\eeq
It is straightforward to show that
\bea
     {\cal E} &=& V_{coul}(R) + E_{se} = \lim_{T\ra 0} V(R,T)
\non \\
     {\cal E}_{min} &=& V(R) + E'_{se} = \lim_{T\ra \infty} V(R,T)
\eea
where ${\cal E}_{min}$ is the minimal energy state containing two
static charges, and $V(R)$ is the static Coulomb potential.  From this
it follows, since ${\cal E}>{\cal E}_{min}$, that if $V(R)$ is
confining, then so is $V_{coul}(R)$, with $V(R) \le V_{coul}(R)$.
This means that a confining Coulomb potential is a \emph{necessary}
(but not sufficient) condition for confinement, as first noted by
Zwanziger \cite{Dan1}.  With a lattice regularization
\bea
L(x,T) &=& U_0(x,a) U_0(x,2a) \cdot \cdot \cdot U_0(x,T)
\non \\
V(R,0) &=& -{1\over a} \log[G(R,1)]
\eea
This allows us to obtain an estimate, exact in the continuum limit,
of $V_{coul}(R)$, which we may compare to the static potential $V(R)$.

\begin{figure}[t!]
\includegraphics[width=7truecm]{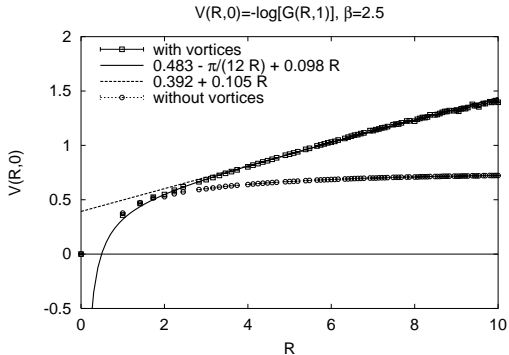}
\caption{$V(R,0)$ at $\b=2.5$, which is a lattice approximation
to the color Coulomb potential.  The solid (dashed) line is a
fit to a linear potential with (without) the L\"uscher term.}
\label{v02p5}
\end{figure}

Figure \ref{v02p5}, upper line of data points, shows our result
for $V(R,0)$ at $\b=2.5$ \cite{GO}. This is clearly a linear
potential; we find however that $\s_{coul} \approx 3\s$. The
evidence, then, is that the color Coulomb potential
\emph{overconfines}, a fact which is relevant to the gluon
chain model, proposed by Greensite and Thorn \cite{gchain}. As $T$
increases, one finds that the slope of $V(R,T)$ approaches $\s$ from 
above, as it should. The
lower line of data points in Fig.\  \ref{v02p5} 
shows the result for $V(R,0)$ when
center vortices are removed from lattice configurations by the 
method of de Forcrand and D'Elia \cite{dFE}. It is clear that removing
center vortices also removes the confining property of the color
Coulomb potential.  In Coulomb gauge, this property is attributed
to the density of near-zero eigenvalues of the Faddeev-Popov
operator; evidently vortex removal must alter this density
drastically.  This question will be addressed in Dan Zwanziger's
talk \cite{Dan_talk}.

\section{The Remnant Symmetry Order Parameter}

    Non-abelian gauge theories can exist in a number of different phases;
those that concern us here belong to one of three broad
categories.  First, there are massless phases, e.g.\ compact
$QED_4$, and non-abelian lattice gauge theories at weak couplings,
and dimension $D>4$, where external charges are associated with
long-range electric fields.  Second, there are confined phases,
e.g.\ SU(N) gauge theory with or without matter fields in the
adjoint representation, in which color electric fields are
squeezed into flux tubes.  In these theories there exists a global
$Z_N$ symmetry of the Lagrangian, which is unbroken at the quantum
level.  Finally, there are screened phases, where the color
electric field falls off exponentially away from the source.  This
condition is found in theories with trivial center symmetry,
such as SU(N) gauge theories with matter in the fundamental
representation, or in $G_2$ pure gauge theory.  The screened phase
is also found when a non-trivial center symmetry is spontaneously
broken, as in high-temperature gauge theory, and in theories
with adjoint matter fields in the Higgs phase.

   Large-scale field fluctuations are enhanced in the confined
phase and suppressed in the screened phases, as can be verified by
computing the appropriate observables.  We are interested in
studying the behavior of the Coulomb energy in these different
phases, and for this purpose it is useful to introduce a new order
parameter.

\begin{figure}[htb]
\includegraphics[width=7truecm]{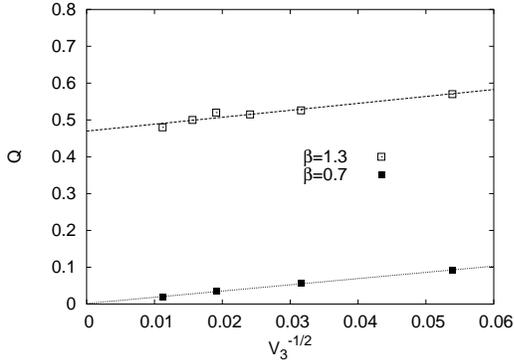}
\caption{Plot of $Q$ vs.\ root inverse 3-volume, and
extrapolation of $Q$ to infinite volume in $QED_4$,
for $\b=0.7$ (confining phase) and $\b=1.3$ (massless phase).}
\label{qv}
\end{figure}

   Minimal Coulomb gauge, which maximizes the average value of
$\sum_{x,k} \mbox{Tr}[U_k(x)]$, does not fix the gauge completely.
There is still freedom to carry out time-dependent
transformations
\bea U_k(x,t) &\ra& g(t) U_k(x,t) g^\dg(t)
\non \\
U_0(x,t) &\ra& g(t) U_0(x,t) g^\dg(t+1)
\eea
which we will call the \emph{remnant symmetry} of Coulomb gauge.
Remnant symmetry is a global symmetry at fixed time; and it
can therefore be
spontaneously broken at a fixed time $t$ in an infinite spatial
volume.  If remnant symmetry is broken, this means that
\beq
\langle U_0(x,t) \rangle \ne 0
\eeq
and in consequence
\bea
\lim_{R\ra \infty} G(R,1) &>& 0
\non  \\
\lim_{R\ra \infty} V(R,0) &=& \mbox{finite const.}
\non \\
      \s_{coul} &=& 0
\label{consequence}
\eea
So Coulomb confinement or non-confinement can be understood as the
symmetric or broken realization, respectively, of the remnant gauge
symmetry (this is not a new idea, c.f.\ ref.\ \cite{Marinari}).
Our order parameter for remnant symmetry breaking, denoted $Q$, is the
modulus of the average timelike link variable, averaged over all sites
at fixed time, i.e.
\bea
U^{av}_0(t) &=& {1\over L^3} \sum_{\vec{x}} U_0(\vec{x},t)
\non \\
Q &=& \left\langle \sqrt{\oh \mbox{Tr}[U_0^{av}(t) U_0^{av \dg}(t)]}
\right\rangle
\eea
On general grounds, on an $L^3 \times L_t$ lattice,
\beq
Q = c + {b \over L^{3/2}}
\eeq
where $c=0$ in the symmetric phase, and $c>0$ in the broken phase.
$Q>0$ at infinite $L$ implies that $V_{coul}(R)$ is non-confining,
and therefore $Q=0$ is a necessary (but not sufficient) condition
for confinement.

\section{Q in Different Phases}

   It is interesting to apply this order parameter first in compact
$QED_4$, where we know there is a transition from the confined to the
massless phase around $\b=1.0$.  Fig.\ \ref{qv} shows our result for $Q$ at
couplings in the confining and massless phases. The values for $Q$
extrapolate to zero at infinite volume in the confined phase, and to a
non-zero value in the massless phase, as expected.

   Next we have computed $Q$ in an SU(2) gauge theory with a scalar field,
of fixed modulus $|\phi|=1$, in the adjoint representation
\bea
   &S& = \b \sum_{plaq} \oh \mbox{Tr}[UUU^\dg U^\dg] 
\\
 &+& {\gamma \over 4} \sum_{x,\m} \phi^a(x)\phi^b(x+\widehat{\m})
       \oh \mbox{Tr}[\s^a U_\m(x) \s^b U_\m^\dg(x)]
\non
\eea
In this case it is also well known that there are two distinct phases in
the $\b - \gamma$ coupling plane, namely,
a confining phase and a Higgs phase, and in this case again we find that
$Q=0$ in the confined phase, and $Q>0$ in the Higgs phase, upon extrapolation
to infinite volume.

\begin{figure}
\includegraphics[width=7truecm]{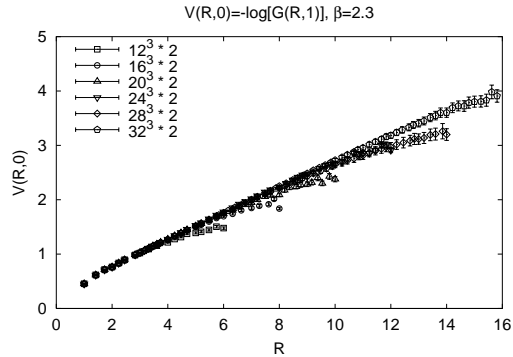}
\caption{$V(R,0)$ in the deconfined phase, at $\b=2.3$
with $n_t=2$ lattice spacings in the time direction, and space
volumes ranging from $12^3$ to $32^3$.}
\label{v1}
\end{figure}

\begin{figure}
\includegraphics[width=7truecm]{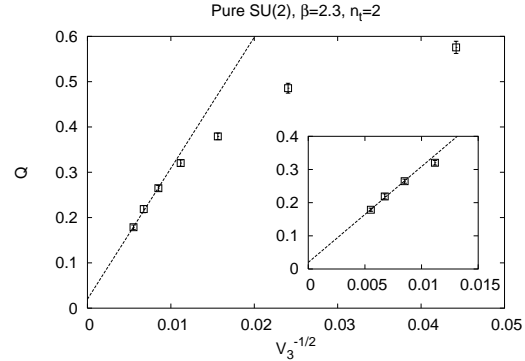}
\caption{The $Q$ parameter vs.\ root inverse 3-volume in the
high-temperature deconfined phase, pure SU(2) gauge theory at
$\b=2.3$ and $n_t=2$ lattice spacings.}
\label{q1}
\end{figure}

   At finite temperature, in pure SU(2) gauge theory, we encounter something
unanticipated.  We have calculated $V_{coul}(R)$ and $Q$ in the
high-temperature deconfined phase,
expecting to see $\s_{coul}=0$.  Instead, the opposite result was obtained,
as seen in Figs.\ \ref{v1} and \ref{q1}.  This result is, however, a little
less surprising when we remember that the non-local, Coulomb part of the
Hamiltonian depends only on links in the three-space directions, and not on
timelike links.  The links in the space directions still form a confining
ensemble, since spacelike Wilson loops have an area law even in the deconfined
phase.  From this point of view, it is rather natural that the instantaneous
Coulomb potential is confining, even if the static quark potential is not.

   As a check of this reasoning, we remove center vortices from the 
high-temperature
ensemble of lattices; this is known to remove the area law for spacelike Wilson
loops, so the spacelike links at fixed time are no longer a confining ensemble.
On computing the color Coulomb potential, we indeed find that $\s_{coul}=0$
asymptotically.

   Finally, we have computed $Q$ in SU(2) gauge theory with a radially-frozen
Higgs field in the fundamental representation.  For the SU(2) gauge group, the
action can be written
\bea
   S &=& \b \sum_{plaq} \oh \mbox{Tr}[UUU^\dg U^\dg]
\non \\
 &+& \gamma \sum_{x,\m} \oh
         \mbox{Tr}[\phi^\dg(x) U_\m(x) \phi(x+\widehat{\m})]
\eea
where $\phi$ is SU(2)-group valued, and $\phi \phi^\dg = I$.  In this
theory the Wilson loop never aquires an area law, at any $\gamma > 0$.
There is a theorem by Fradkin and Shenker \cite{FS} which says that it
is always possible to follow a path from the small-$\g$
"confinement-like" region of the $\b-\g$ phase diagram, into the
large-$\g$ Higgs region, which avoids any non-analyticity in local
gauge-invariant observables, such as the free energy.  Thus, according
to the textbook definition, the Higgs and confinement-like regions do
not constitute distinct phases of the lattice theory.

\begin{figure}[htb]
\includegraphics[width=7truecm]{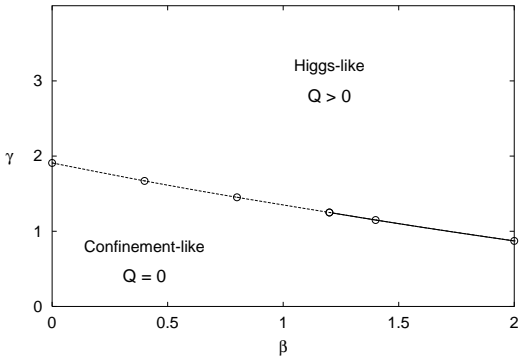}
\caption{Phase diagram of the SU(2) fundamental Higgs model.
There is a thermodynamic transition and a $Q$ (remnant
symmetry-breaking) transition along the solid line, but only
a non-thermodynamic transition
(Kert\'esz line) along the dashed line.}
\label{phase_f}
\end{figure}

\begin{figure}[htb]
\includegraphics[width=7truecm]{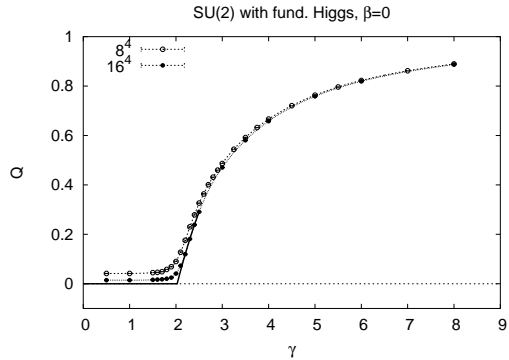}
\caption{$Q$ vs.\ $\g$ at $\b=0$ in the SU(2) fundamental Higgs
model, on $8^4$ and $16^4$ lattices.  The solid line is the
presumed extrapolation of $Q$ to infinite volume.}
\label{qbeta0}
\end{figure}

   The $Q$ operator tells a different story.  We find, in fact, two
distinct regions: a region of unbroken remnant symmetry, where $Q=0$,
and a region of broken symmetry, with $Q>0$, as shown in Fig.\
\ref{phase_f}.  The boundary between these regions is formed by the
solid line, which is a line of first-order phase transitions, continuing
into the dashed line, which is not associated with any
non-analyticity in the free energy.  There is a discontinuity in $Q$
across the solid line.  $Q$ is zero at the dashed line, and then rises
to non-zero values as $\g$ increases, as illustrated in Fig.\
\ref{qbeta0}. This behavior is strongly reminiscent of magnetization
in a second order phase transition.

    The non-analytic behavior of the $Q$ order parameter, across the
remnant-symmetry breaking transition, is not at odds with the
Fradkin-Shenker theorem.  When expressed in gauge-invariant form, $Q$
is a highly non-local observable, whose behavior is not covered by
this theorem.  Nevertheless, we would like to better understand the
nature of a symmetry-breaking transition which is not accompanied by
any singularity in thermodynamic quantities.  It was suggested by
Langfeld \cite{Kurt1} that this is an example of a Kert\'esz line
\cite{Kertesz}.

   A Kert\'esz line is a line of percolation transitions; the original
example comes from the Ising model.  In the Ising model, in the
absence of an external magnetic field, there is a phase transition
from a $Z_2$ symmetric phase to an ordered phase. This transition can be
expressed, in different variables, as a transition from a percolating
phase at low temperature, to a non-percolating phase at high
temperature.  In the presence of a magnetic field, the partition
function and thermodynamic observables become analytic in temperature;
there is no thermodynamic phase transition.  Nevertheless, the
percolation transition persists, and traces out a Kert\'esz line in
the temperature-magnetic field plane, completely separating the phase
diagram into two regions.  The interesting question, in the
gauge-Higgs theory, is what sort of objects are actually percolating.
Based in part on results reported by Bertle and Faber \cite{Bertle1},
Langfeld \cite{Kurt2} conjectured that the unbroken remnant symmetry
region is a region of percolating center vortices, which cease
percolating in the broken symmetry region.

   This conjecture that the gauge-Higgs theory is divided into
percolating and non-percolating phases has been verified in ref.\
\cite{Bertle2}, which introduces a percolation observable $s_w$.  To
define this observable, let $f_p$ denote the fraction of all
P-plaquettes on the lattice, which lie in the P-vortex containing the
P-plaquette $p$.  Then $s_w$ denotes the value of $f_p$ averaged over
all P-plaquettes.  This is, in a sense, the fraction of the total
P-vortex area contained in the "average" P-vortex.  If all
P-plaquettes on the lattice lie a single percolating vortex, then
$s_w=1$.  If vortices do not percolate at all, then $s_w=0$ on an
infinite lattice.  We have percolation, in infinite volumes, when
$s_w>0$.

\begin{figure}[htb]
\includegraphics[width=7truecm]{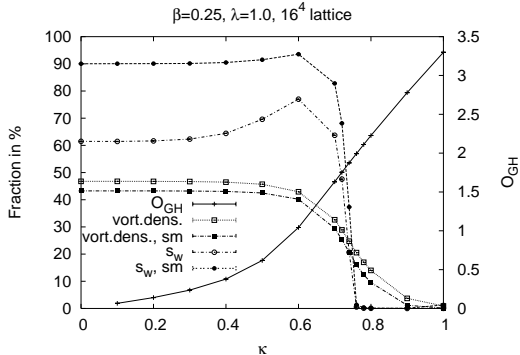}
\caption{Vortex percolation order parameter $s_w$ (open circles)
vs.\ gauge-Higgs coupling $\kappa$, in the gauge-Higgs theory of eq.\
\rf{ghiggs}.  The other couplings ($\b=0.25,~\l=1.0$) are fixed.  Also
shown is the vortex density, and the one-link observable $O_{GH}$.
The "sm" data points refer to a type of smoothing, as explained in
ref.\ \cite{Bertle2}.}
\label{bertle}
\end{figure}

   Ref.\ \cite{Bertle2} did not actually use a radially frozen Higgs
field; the matter action of the gauge-Higgs theory was instead taken
to be
\bea
  \lefteqn{S_{matter} =} & &
\non \\
    & &\sum_x \left[\Phi^\dagger(x)\Phi(x) +
    \lambda\left(\Phi^\dagger(x)\Phi(x)-1\right)^2\right]
\non \\
    & & - \kappa \sum_{\mu,x} \left(\Phi^\dagger(x)U_\mu(x)\Phi(x+\hat{\mu}) +
    \mbox{c.c.}\right)
\label{ghiggs}
\eea
where $\Phi$ is a two-component massive scalar field in the
fundamental representation of SU(2).  The result for $s_w$ vs.\
$\kappa$ at $\b=0.25,~\l=1.0$ is shown in Fig.\ \ref{bertle}, along
with a number of other observables such as vortex density and
$O_{GH}=\langle \Phi^\dagger(x)U_\mu(x)\Phi(x+\hat{\mu}) \rangle$ .
In local gauge-invariant observables, such as $O_{GH}$,
there is no obvious transition as $\kappa$ increases.  But in the center
vortex percolation variable $s_w$, a sharp transition is seen.

  Our findings in the gauge-Higgs theory bear on the question: In what
sense does real QCD, or any gauge theory with matter in the
fundamental representation, confine color?  As already noted, in such
theories there are no transitions in the free energy which isolate the
Higgs from the confinement-like regions of the phase diagram; it
\emph{seems} possible to continuously interpolate from one region to
another.

   In contrast, we find that these phases are indeed physically
different, and separated by a sharp symmetry-breaking transition.  The
confinement-like phase is distinguished from the Higgs phase by its
unbroken realization of remnant symmetry, by a confining color Coulomb
potential, and by percolating center vortices.

\section{Conclusions}

    We have found that the color Coulomb potential in pure SU(2) gauge
theory is linearly rising, with a slope which is roughly three times
larger than the usual asymptotic string tension.  This overconfinement
is essential to the gluon-chain scenario of QCD string formation.

     The confinement property of the color Coulomb potential is a
consequence of the unbroken realization of remnant gauge symmetry in
Coulomb gauge.  Center symmetry breaking, which takes $\s \ra 0$, does
not \emph{necessarily} imply remnant symmetry breaking.  In at least
two cases, namely, the high-temperature deconfined phase and the
confinement-like phase of gauge-fundamental Higgs theory, we find that
$\s=0$ coexists with $\s_{coul}>0$.  In the latter theory, the
transition to the Higgs phase is a remnant-symmetry breaking, center
vortex depercolation transition.

   In every case, center vortex removal also sends $\s_{coul}\ra 0$,
suggesting that there may be a deep relationship between the center
vortex and Gribov horizon confinement scenarios.  This relationship
will be explored more fully in Dan Zwanziger's talk \cite{Dan_talk} at
this meeting, and in ref.\ \cite{new}.

\end{document}